\documentclass[aps, prl, citeautoscript, amsmath, amssymb, reprint]{revtex4-1}
\usepackage{graphicx}
\usepackage{color}
\usepackage{cases,array}
\usepackage{bm}

\usepackage[english]{babel}

\usepackage{hyperref}

\usepackage{commath}

\begin{document}

\title{Nonlinear behavior and mode coupling in spin transfer nano-oscillators}

\author{R. Lebrun}
\affiliation{Unit\'e Mixte de Physique CNRS/Thales, Palaiseau and Universit\'e Paris-Sud, Orsay, France}

\author{N. Locatelli}
\affiliation{Unit\'e Mixte de Physique CNRS/Thales, Palaiseau and Universit\'e Paris-Sud, Orsay, France}

\author{F. Abreu Araujo}
\affiliation{Institute of Condensed Matter and Nanosciences, Universit\'{e} catholique de Louvain, Louvain-la-Neuve, Belgium}

\author{H. Kubota}
\author{S. Tsunegi}
\author{K. Yakushiji}
\author{A. Fukushima}
\affiliation{National Institute of Advanced Industrial Science and Technology (AIST), Spintronics Research Center, Tsukuba, Japan}

\author{J. Grollier}
\affiliation{Unit\'e Mixte de Physique CNRS/Thales, Palaiseau and Universit\'e Paris-Sud, Orsay, France}

\author{S. Yuasa}
\affiliation{National Institute of Advanced Industrial Science and Technology (AIST), Spintronics Research Center, Tsukuba, Japan}

\author{V. Cros}
\affiliation{Unit\'e Mixte de Physique CNRS/Thales, Palaiseau and Universit\'e Paris-Sud, Orsay, France}
\altaffiliation{corresponding author : vincent.cros@thalesgroup.com}

\begin{abstract}
By investigating thoroughly the tunable behavior of coupled modes, we highlight how it provides new means to handle the properties of spin transfer nano-oscillators. We first demonstrate that the main features of the microwave signal associated with coupled vortex dynamics i.e. frequency, spectral coherence, critical current, mode localization, depends drastically on the relative vortex core polarities. Secondly we report a large reduction of the nonlinear linewidth broadening obtained by changing the effective damping through the control of the core configuration. Such a level of control on the nonlinear behavior reinforces our choice to exploit the microwave properties of collective modes for applications of spintronic devices in novel generation of integrated telecommunication devices.
\end{abstract}

\maketitle

Spin transfer torque has revealed the potential of spintronic devices for a new generation of electronic components showing multiple functionalities \cite{Locatelli2014, Stiles2006}, notably for microwave applications \cite{Slavin2009}. Within this palette of novel applications, spin-transfer nano-oscillators (STNOs) relying on the conversion of nonlinear magnetization dynamics into a microwave signal is anticipated as being the most promising one, given their high spectral coherence \cite{Kim2008, Keller2010}, large tunability with current \cite{Slavin2009}, frequency modulation properties \cite{Martin2013, Iacocca2012} and ability to synchronize to an external signal \cite{Georges2008_2, Demidov2014}. Indeed these theoretical and experimental studies have emphasized the importance of their nonlinear character on their microwave properties. More recently,  several studies have also revealed the potential strong influence of mode coupling on the nonlinearities of STNO \cite{Sankey2005, Iacocca2014, Krivorotov2008, Deac2008, Cherepov2012, Pulecio2014}. A direct consequence is that mode coupling should now be considered as a strategy to tune their intrinsic nonlinearities  \cite{Gusakova2011}. A reduction of nonlinearities can lead to a cancellation of the undesired linewidth broadening, but concurrently to a decrease of the frequency tunability with current. However contrary to uniformly magnetized STNOs it has been highlighted that the vortex case have the ability to conserve large frequency tunability with current through the Oersted field influence \cite{Pribiag2007}.

In this article, we aim at emphasizing the interest of spin transfer oscillators based on the current induced dynamics of two weakly coupled vortices. In such system, spin transfer will give rise to self-sustained oscillations of hybridized gyrotropic modes, each of them depending on the relative configuration of each vortex (core polarity and chirality).
Combining experiments, analytics and numerical simulations we investigate the influence of relative vortex core configurations on the properties of coupled modes. We report not only the effect of core polarities on mode frequency and gyration radii but also on the evolution of the critical current and the linewidth broadening through a modification of the non-linear parameters. Notably we demonstrate that a strong reduction of the non-linearity through an increase of the effective damping term can be achieved by choosing properly the excited coupled mode.  Thus, coupled vortices appears to be a model system \cite{Locatelli2011} for the study and the improvement of the properties of spin transfer nano-oscillators through collective modes dynamics.\\

\begin{figure}[ht!]
\includegraphics[scale=.45]{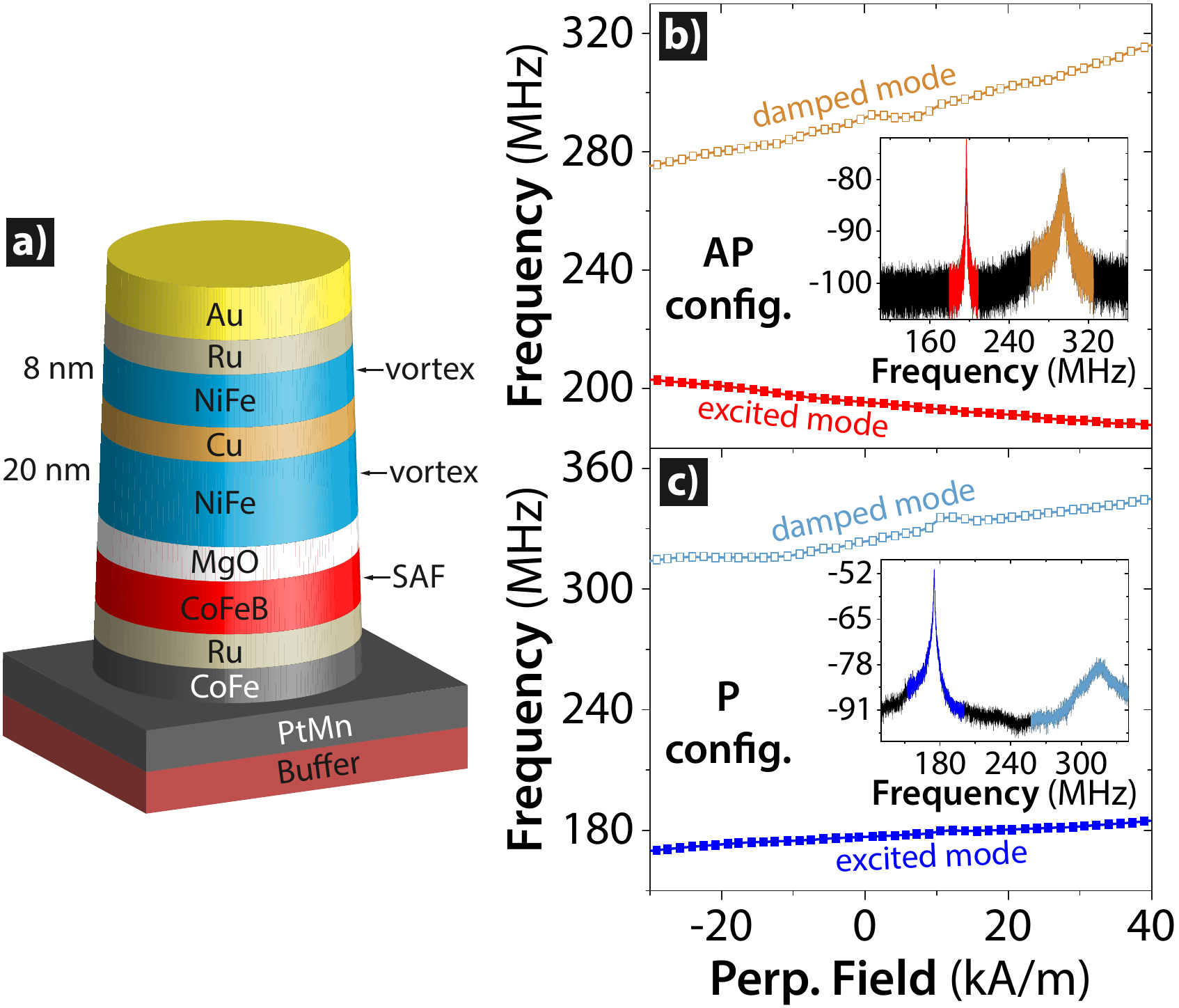}
\caption{(Color online) (a) Schematic of an hybrid magnetic tunnel junction: a Cu based spin-valve system with the two vortex Py layer (8 nm at the top, 20 nm at the bottom) above a 1 nm MgO barrier and a CoFeB based synthetic antiferromagnet. (b) \& (c) Field dependence of the low frequency "excited" mode and the high frequency "damped" mode for AP (b) and P (c) configuration at $I_\text{dc}=-16$ mA. Inset: Frequency spectrum of  the output emitted signal at $H_\text{applied} = 0$ kA/m for AP (b) and P (c) configuration.}
\label{fig1}
\end{figure}

The studied samples are prototype nanopillars with a nominal 300 nm radius made from a multilayer stack containing a CoFe/Ru/CoFeB synthetic antiferromagnet (SAF), a MgO barrier and then a NiFe(20nm)/Cu/NiFe(8nm) spin-valve (Fig. \ref{fig1}c). Each NiFe layer has a vortex magnetic configuration. Given that the total thickness is much larger than for standard MTJ, the etching process during the nanofabrication results in a conic shape pillar with a 290 nm radius for the top thin NiFe layer and 340 nm radius for the bottom thick layer.

In such configuration with 2-vortices (2V), the existence of  dipolar coupling between the two vortices implies that the two gyrotropic modes associated to each vortex will hybridize \cite{Guslienko2002}. Each of these two coupled modes being predominantly associated to one of the vortices, the one that will be effectively excited by spin torque depends on the sign of injected current. In our system, every combination of chiralities and core polarities can be obtained by careful magnetic preparation. However, in the following, we will restrain our investigation on the configuration with identical vortex chiralities, being parallel to the Oersted field. For this case, we will compare the spin transfer dynamics of the coupled modes measured for identical (P) and opposite (AP) polarities configurations under a negative dc current (electrons flowing from the top thin layer vortex to the bottom thick layer vortex), and compare the dynamics. It is to be noticed that because the TMR ratio is about 70\% compared to a GMR ratio of 3\% in the spin-valve part, we essentially detect the motion of the coupled modes (and thus the emitted power) only through the vortex dynamics in the 20 nm thick NiFe layer that is in contact with the MgO barrier.

For this configuration, we anticipate the spin transfer torque will damp the coupled gyrotropic motion mainly driven by the thick layer vortex but excite the one driven by the thin layer vortex \cite{Khvalkovskiy2010_2}\footnote{These measurements are performed at zero or small applied perpendicular magnetic field, allowing us to neglect additional contributions to spin torque arising from the SAF of the MTJ \cite{Khvalkovskiy2010_2}}, that has a lower frequency. Thus, for both relative configurations, we will generate spin transfer oscillations of the coupled vortex system through the hybridized mode dominated by the thin layer.

Indeed, as shown in the insets of Fig. \ref{fig1} (b,c), we detect for both cases a peak having a large amplitude and a narrow linewidth at a frequency that is close to the predicted gyrotropic frequency for the isolated thin layer’s vortex (197 MHz). We also record a much broader peak (linewidth above 4 MHz) at higher frequency which is attributed to thermal excitation of the second coupled mode. In Fig. \ref{fig1}c (resp. Fig. \ref{fig1}b) we show for parallel (P) (resp. anti-parallel (AP)) cores polarities the frequency evolution of these two modes as a function of the perpendicular applied field $H_\text{perp}$. As expected \cite{Deloubens2009}, the slopes of these modes have the same (resp. opposite) sign for  parallel (resp. anti-parallel) core polarities. We also note that the best spectral coherence for the excited mode has been measured in the AP configuration, with a minimum linewidth of 80 kHz (at $H_\text{perp} = 30$ kA/m and $I_\text{dc} = - 16$ mA) leading to a Q factor of 2400 compared to a $Q_\text{max}=300$ in P configuration.

\begin{figure}[ht!]
\includegraphics[scale=.43]{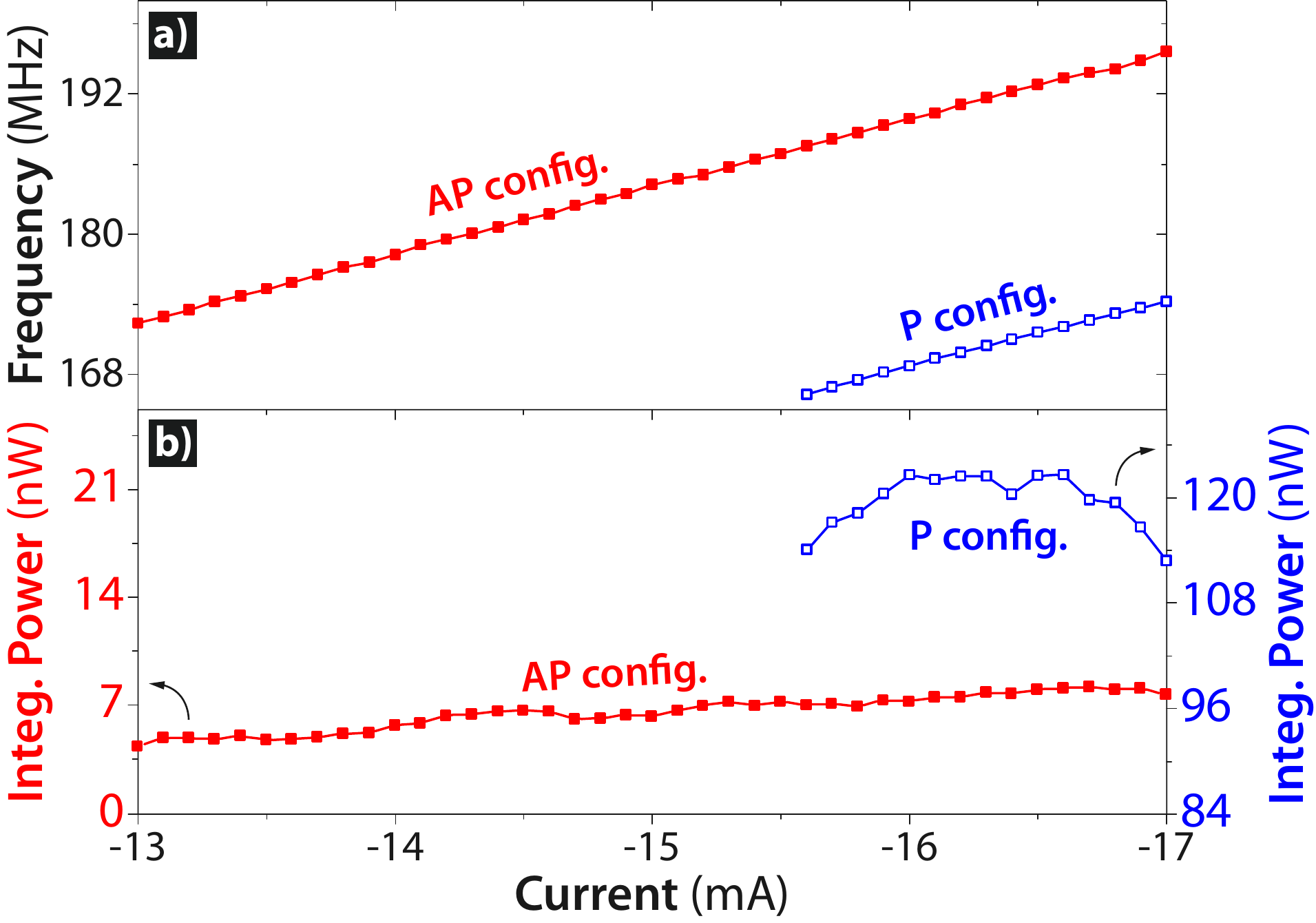}
\caption{(Color online) Frequency (a) and integrated output emitted power (b) current dependency for AP (red dot) and P (blue dot) configurations at zero applied field.}
\label{fig2}
\end{figure}

In Fig. \ref{fig2}(a,b), we present the evolution with the dc-current $I_\text{dc}$ of both frequency and integrated power of the excited low frequency mode for both P and AP cores configurations at zero applied magnetic field. Several important features can be noticed. First, we find an almost strictly linear dependence of the frequency with $I_\text{dc}$, a feature that is interesting for frequency modulation using STNOs. Moreover the two $df/dI_\text{dc}$ slopes are identical and are separated by a constant frequency difference of 20 MHz. Second, the output emitted power of the excited mode is about 10 times smaller in AP core configuration (10 nW) compared to P case (100 nW). Third, the threshold current $I_\text{c}$ is found to be much lower in the AP configuration ($I_\text{c-AP}=-12$ mA and $I_\text{c-P}=-15.6$ mA).

To elucidate these large differences of microwave features of the excited modes in the two cores configurations, we developed an analytical model based on the Thiele formalism \cite{Guslienko2005}. In order to describe our system, we add a coupling term that accounts for the dipolar interaction between the two in-plane mean magnetizations of the moving vortices (so-called body-body interaction). The dipolar core-core interactions are neglected, given that the two cores are far from each other when they are moving under the action of the spin torque. In complex coordinates $\bm{X} \equiv X \text{e}^{i\theta}$, we obtain the following system of coupled equations:
\[
\left\lbrace
\begin{array}{l}
\displaystyle \frac{d\bm{X}_1}{dt} - \frac{k_1(J_1)}{(ip_1G_1 - D_1)}\bm{X}_1 - \frac{\mu}{(ip_1G_1 - D_1)}\bm{X}_2 = 0\\
\displaystyle \frac{d\bm{X}_2}{dt} - \frac{k_2(J_2)}{(ip_2G_2 - D_2)}\bm{X}_2 - \frac{\mu}{(ip_2G_2 - D_2)}\bm{X}_1 = 0
\end{array}
\right.
\]

with $k_{1,2}$ (J) the magnetostatic and Oersted field confinement coefficient \cite{Dussaux2012}, $p_{1,2}$ the vortex core polarity, $J_{1,2}$ the current density, $G_{1,2}$ the gyrovector, $D_{1,2}$ the dyadic damping term  \cite{Dussaux2012}, and $\mu$ the dipolar body-body coupling term. The index 1 (resp. 2) stands for the thick (resp. thin) layer vortex.
By analytically solving the system \footnote{We use the parameters (thickness and diameter) corresponding to our systems with a NiFe magnetization, $M_\text{s} = 600$ emu.cm$^{-3}$ and $\alpha=0.01$}, we obtain for each relative configuration two eigenvalues $\lambda_{a,b}$ and two eigenvectors $\bm{V_{a,b}}$, and so we can extract the hybridized resonant mode frequencies Im($\lambda_{a,b}$) as well as the ratio of gyration radii in each layer $(\rho_{1} / \rho_{2})_{a,b}$ = $(V_{a,b})_{X_1}$ / $(V_{a,b})_{X_2}$ with $(V_{a,b})_{X_{1,2}}$ the projection of the eigenvectors in the basis ($\bm{X}_1$,$\bm{X}_2$). Note that, at this stage, neither the spin transfer torque nor the nonlinear contribution of the confining force and the damping force are taken into account.

\begin{figure}[ht!]
\includegraphics[scale=.45]{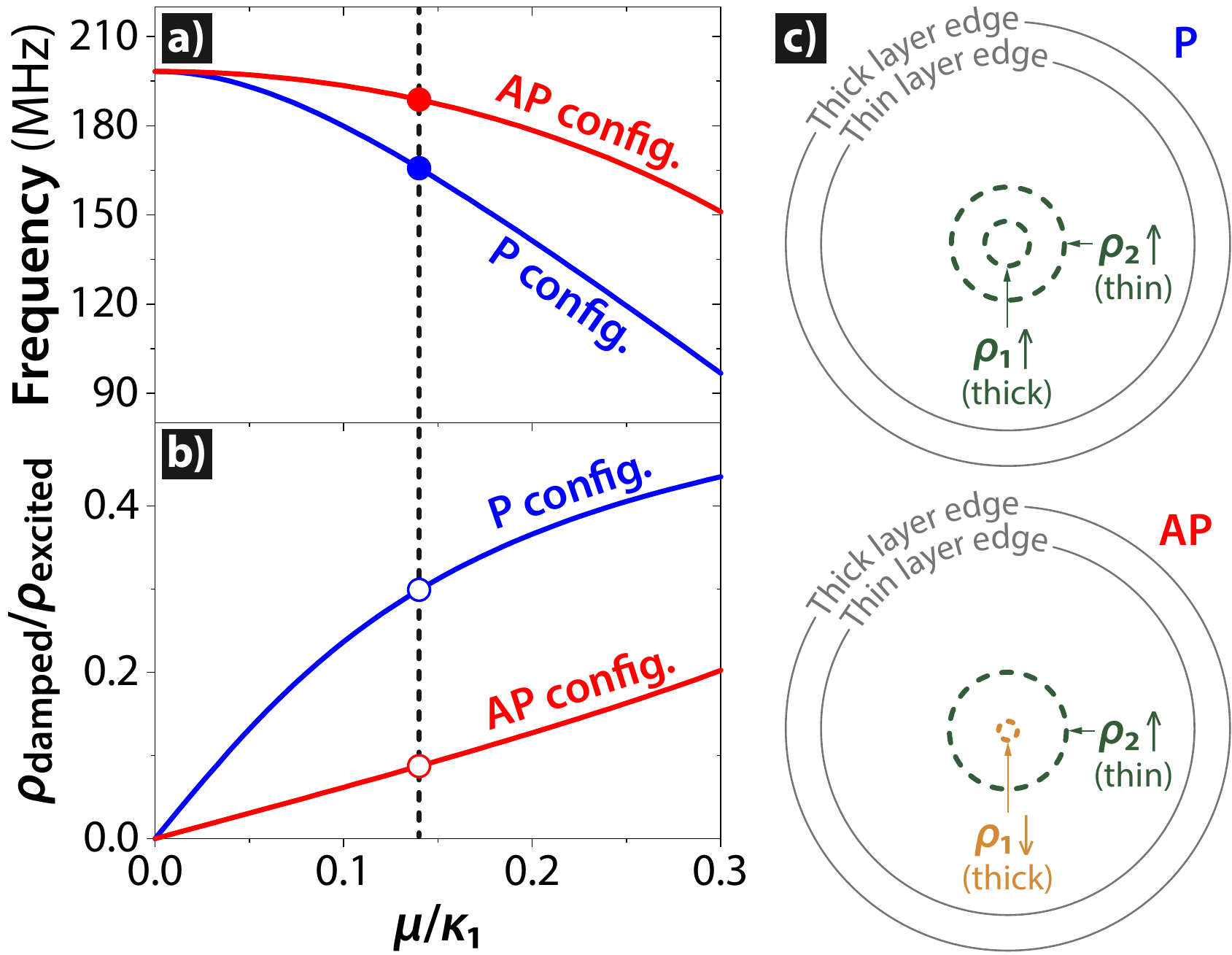}
\caption{(Color online) Frequency vs coupling for the lowest excited mode in parallel core (blue) and antiparallel core (red) for $I_\text{dc} = -16$ mA at zero field (b) Ratio of gyration radii between thin and thick layer for the lowest frequency mode in parallel (blue) and anti-parallel (red) configurations [Filled (unfilled) circles represents the (expected) experimental points] (c) Micromagnetic simulations representing gyration radiuses $(\rho_{1},\rho_{2})$ for P and AP configurations at zero field and $I_\text{dc} = -16$ mA. Dot edges (black lines) present a 50 nm difference due to ion etching (see Fig. 1 (a))}
\label{fig3}
\end{figure}

In Fig. \ref{fig3}, we display the evolution of the frequency (Fig. \ref{fig3}a) and the radii (Fig. \ref{fig3}b) of the low frequency mode as a function of the coupling strength normalized to the confinement $\mu / \kappa_1$ for P core configuration (blue line) and for the AP one (red line). These analytical predictions have been compared to micromagnetic simulations including the spin transfer torque as well as the Oersted field and we obtain an excellent overall agreement.

\begin{figure*}[ht!]
\includegraphics[scale=.9]{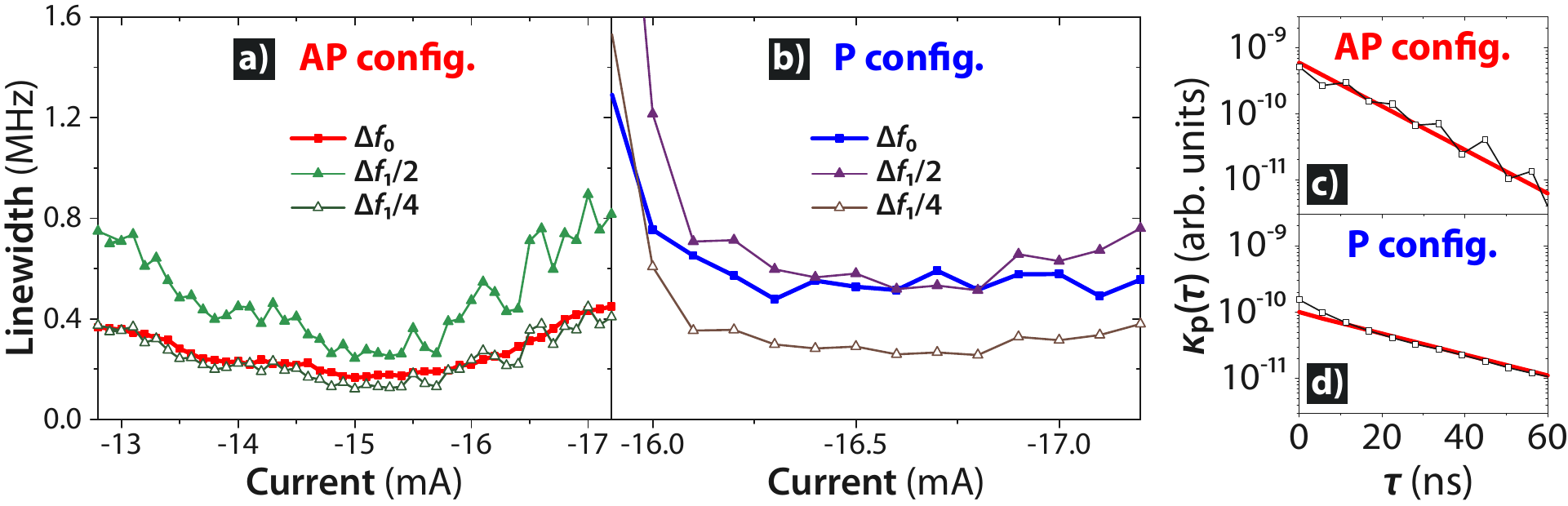}
\caption{(Color online) Spectral linewidth of the fundamental and the first harmonics divided by two (filled symbols) and four (open symbols) for AP (a) and P (b) configuration measured at zero applied magnetic field. Autocorrelation function of power fluctuations for both configurations AP (c) and P (d) at zero field for $I_\text{dc} = -17$ mA.}
\label{fig4}
\end{figure*}

From the 20 MHz frequency difference between the two cores configurations found experimentally we can estimate the matching value of the coupling coefficient (see dotted line in Fig \ref{fig3}a and b to be equal $\mu = 0.14 \kappa_1$, as well as the corresponding gyration radii ratio from Fig \ref{fig4}b: $\rho_1 / \rho_2 = 0.3$ for identical polarities and $\rho_1 / \rho_2 = 0.1$ for opposite polarities. These latter predictions for the ratio of the gyrotropic radius for each vortex has been confirmed by micromagnetic simulations done at $I_\text{dc} = -16$ mA (see Fig. \ref{fig3}c). Indeed, we find similar radii for the gyrotropic motion of the thin layer vortex (about 90 nm) for both core configurations whereas the radius of the thick layer vortex motion strongly depends on the core configuration:  9 nm in AP and 32 nm in P. Finally, we would like to emphasize that, even when vortices have opposite core polarities, the two vortex cores are gyrating in the same direction, thus confirming that the spin transfer torque do excite a single coupled mode and not two independent gyrotropic motions.

Indeed the difference of gyration amplitude of the thin and thick vortex layer shown in Fig. \ref{fig3}b allows explaining, even quantitatively the large difference of emitted power found experimentally between P and AP core configuration (see Fig. \ref{fig2}b) : 100nW/10nW $\propto (\rho_{1,P} / \rho_{1,AP})^2$ \cite{Slavin2009}. In fact, this strong difference of amplitude motion in the thick layer cannot be attributed to the influence of the spin transfer (as it is not taken into account in the analytical model) but rather to the sense of gyration of the vortex thick layer that is opposite to its “normal” one in AP configuration \cite{Guslienko2002}.\\

To tackle the issue of the observed different spectral coherences related to the dynamical properties of these coupled modes, it is anticipated that the non-linear character of spin transfer oscillators should play a crucial role. Indeed several approaches have been proposed to evaluate the influence of non-linearities on the STNO linewidth \cite{Kim2008, Iacocca2014, Rowlands2013, Lee2013}. Notably, it has been demonstrated that the normalized dimensionless nonlinear frequency shift $\nu$ implies a conversion of amplitude fluctuations into undesired phase fluctuations. It has been proposed recently that this important parameter can be extracted through the analysis of the signal harmonics linewidths \cite{Quinsat2012}.

In Fig. \ref{fig4}, we present the evolution of the linewidth of the low frequency coupled mode and its first harmonics as a function of $I_\text{dc}$. In AP core configuration (Fig. \ref{fig4}a), we find that the ratio between the linewidth of the fundamental mode $\Delta f_0$ and the first harmonic $\Delta f_1$ is close to 4 in the all range of injected current. Such behavior is consistent with an oscillator that is quasi-isochronous ($\nu \ll 1$) for with it has been predicted the linewidth of the nth harmonic is $\Delta f_n / \Delta f_0 = (n+1)^2$. This behavior strongly differs for P configuration (Fig. \ref{fig4}b) for which we find a ratio close to 2 between $\Delta f_0$ and $\Delta f_1$ for all $I_\text{dc}$. This corresponds to a non-isochronous oscillator (large $\nu$), a case for which $\Delta f_n / \Delta f_0 = (n+1)$.

This latter case obtained for P core configuration appears to be similar to the single vortex case in which a large non-linear frequency shifts $\nu$ have been found \cite{Grimaldi2014, Sanches2014}. Moreover, the strong reduction of nonlinearity in AP configuration is consistent with the much smaller linewidth (about 100 kHz) we found in AP configuration because of the absence of linewidth broadening from non linearities. At last, we emphasize that these features have been reproduced for several other samples and are also consistent with our previous studies in spin-valve nanopillars for AP configuration \cite{Hamadeh2013}.

The normalized dimensionless nonlinear frequency shift $\nu$ is expressed as $\nu = Np/\Gamma_\text{p}$ where $N$ is the nonlinear frequency shift, $p$ is the normalized oscillation power (note this power comes from the motion of both thick and thin layer vortex) and $\Gamma_\text{p}$ is the effective damping rate that describes how fast an oscillator returns to its stable trajectory after a deviation of its amplitude. The parameter $N$ can be estimated experimentally from the evolution of the frequency with $I_\text{dc}$ : $f = f_0(I) + 2\pi N p^2$. The fact that we found similar $df/dI_\text{dc}$ for both core configurations suggests that the nonlinear frequency shift $N$ is indeed small and in any case of comparable amplitude. Then the normalized power p can be easily estimated from micromagnetic simulations. Indeed, we estimate than the normalized oscillation power is slightly higher in P configuration (about 12\% higher) than in AP case. This result is due to the fact that, in both cases the gyration radius in the excited thin layer is much larger than in the thick one and is around 90 nm.

The last term entering in the expression of nonlinear frequency shift $\nu$ is the effective damping rate $\Gamma_\text{p}$. This parameter can be extracted from the experiments through the study of the temporal evolution of the signal. To do so, we have performed Hilbert transforms on 5 ms time traces and fitted the autocorrelation function of power fluctuations (as shown on Fig \ref{fig4} c and d) with the following expression \cite{Bianchini2010}:
\[
\kappa_\text{p} = \left\langle \delta p(\tau) \mid \delta p(0) \right\rangle = A(p_0, \Gamma_\text{p} \text{e}^{-2\Gamma_\text{p}\left| \tau \right|})
\]
Through such analysis, we can determine that $\Gamma_\text{p}$ is twice larger in AP configuration ($\sim 30$ MHz) than in P case ($< 13$ MHz). Finally, taking into account all the contributions, we find the nonlinear frequency shift $\nu$ is about 3 times larger in the P core configuration than in AP. It is however to be emphasized that this "small" difference leads to drastically different microwave features, thus demonstrating the importance to finely tune the nonlinear parameters in these spin transfer oscillators. An interesting feature of our coupled vortex oscillators is the factor 2 difference on the effective damping rate $\Gamma_\text{p}$ that could be of a great interest for rf-applications as it is directly linked to the modulation bandwidth \cite{Slavin2009}.

To conclude, we have studied the large influence of vortex cores configurations on the dynamics of the collective modes of the oscillator (frequency, spectral-coherence, critical current). In particular, we have shown the strong correlation between the vortex core configuration and the nonlinear frequency shift of the excited mode $\nu$, a crucial parameter for describing the main rf-features of the microwave signal. Indeed, we demonstrated that the significant reduction of the linewidth broadening due to nonlinearities that is observed for antiparallel core polarities is due to an increase of the effective damping parameter  $\Gamma_\text{p}$. These results highlight the potential of coupled modes for potential radio-frequency, storage, or associative memories applications.\\

The authors acknowledges E.Grimaldi and A.Jenkins for fruitful discussions, Y. Nagamine, H. Maehara, and K. Tsunekawa of CANON ANELVA for preparing the MTJ films and the financial support from ANR agency (SPINNOVA ANR-11-NANO-0016) and EU FP7 grant (MOSAIC No. ICT-FP7- n.317950). F.A.A. acknowledges the Research Science Foundation of Belgium (FRS-FNRS) for financial support (FRIA grant).

\email{V. Cros}
\bibliography{../MyLib.bib}

\end{document}